\documentclass[review]{elsarticle}

\usepackage{lineno,hyperref}
\modulolinenumbers[5]
\usepackage{graphicx,color}
\usepackage{wrapfig}
\usepackage{dcolumn}
\usepackage{gensymb}

\journal{ArXiv}

\begin{document}

\begin{frontmatter}

\title{An experimental analysis from the magnetic interactions in nanowire arrays}

\author{Wibson W. G. Silva$^{1}$, Adrielson de A. Dias$^{2}$, Alexandre R. Rodrigues$^{3}$, Francisco E. Chávez$^{4}$, Rafael A. de Oliveira$^{1, 2}$ and José Holanda$^{1, 2,*}$}
\address{$^{1}$Programa de Pós-Graduação em Engenharia Física, Universidade Federal Rural de Pernambuco, 54518-430, Cabo de Santo Agostinho, Pernambuco, Brazil}
\address{$^{2}$Unidade Acadêmica do Cabo de Santo Agostinho, Universidade Federal Rural de Pernambuco, 54518-430, Cabo de Santo Agostinho, Pernambuco, Brazil}
\address{$^{3}$Departamento de Física, Universidade Federal de Pernambuco, 50670-901 Recife, Pernambuco, Brazil}
\address{$^{4}$Facultad de Biología, Universidad Michoacana de San Nicolas de Hidalgo, Av. F. J. Mujica s/n Cd. Universitaria, Morelia, Michoacán, México}
\cortext[mycorrespondingauthor]{Corresponding author: joseholanda.silvajunior@ufrpe.br}

\begin{abstract}
We study the magnetic interactions experimentally in nanostructures of nanowire arrays. The intensity value obtained from the interactions provides information about its magnetic behavior. We observed two types of experimental magnetic behavior, i. e., demagnetized and magnetized. Our approach represents the first experimental way to analyze the magnetic behavior of a nanostructure considering its magnetic dependence, which is very important for applications in sensors, for example.
\end{abstract}

\begin{keyword}
\texttt{}Energy\sep experimental\sep interactions\sep nanostructures\sep nanowires

\end{keyword}

\end{frontmatter}

\vspace{10cm}
Many research groups investigate the magnetic properties of nanometric systems [1-8]. Such interest is due particularly to the magnetic interactions that influence the properties of nanostructures [9-13]. As it is well known, magnetic interactions lead to complex phenomena that are far from well understood [14-16]. One of the methods used to analyze these effects is the magnetization process. Such a process produces the remanent magnetization used as a source of information about the intensity value from the interactions [17-22]. One can manipulate the magnetic interactions using the remanent state by the $\Delta$m curves. Such $\Delta$m curves are comparisons between isothermal remanent magnetization (IRM(H)) and direct current demagnetization (DCD(H)) curves [23-28].

In this Letter, we present an experimental analysis to determine the intensity value from the magnetic interactions in nanowire arrays electrodeposited on anodic aluminum oxide (AAO) membranes. We electrodeposited the nanowires of different materials (permalloy, iron, and cobalt) and interpreted the experimental results considering the angular dependence from the interactions. Through our analysis we detected two types of magnetic behavior: demagnetized and magnetized. The demagnetized state refers to the process in which the predominant interactions are demagnetizing (PID). On the other hand, the magnetized state is when the predominant interactions are magnetizing (PIM). This type of analysis is of fundamental importance for applications in devices dependent on magnetic behavior [1, 3, 6].

In the \textbf{Fig. 1} we present an illustration of a hexagonal distribution of nanowire arrays electrodeposited in membranes of anodic aluminum oxide (AAO), as in Refs. [3], [4] and [5]. We also define in \textbf{Fig. 1} the $\theta$ and $\theta_{H}$ angles, which are the angles that the magnetization and magnetic field respectively make with the axis of the nanowires during the measurements of IRM(H) and DCD(H) curves, which were performed by using a Vibrating Sample Magnetometer (VSM) model EV7 at room temperature. 
\begin{figure}[h]
	\vspace{0.1mm} \hspace{0.1mm}
	\begin{center}
		\includegraphics[scale=0.25]{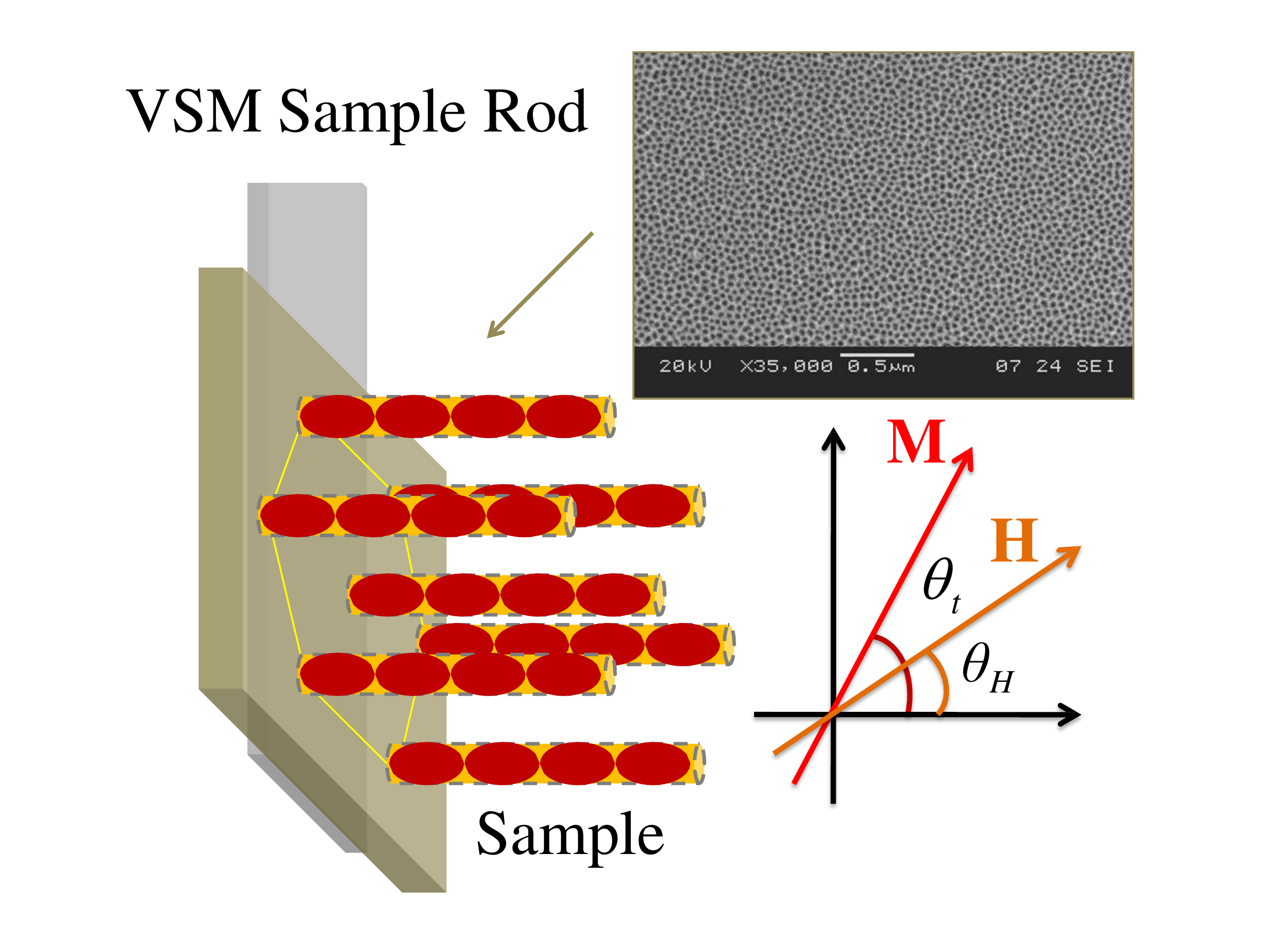}
		\caption{\label{arttype}(Color online) The illustration shows the nanowire arrays and directions of the magnetic field applied during the measurements. The $\theta$ and $\theta_{H}$ angles represent the angles that the magnetization and magnetic field, respectively make with the axis of the nanowires and \textit{t} = PID or PIM.}
		\label{puga}
	\end{center}
\end{figure}
The polycrystalline nanowire (permalloy, iron, or cobalt) arrays were electrodeposited onto AAO membranes. We manufacture some samples, i. e, eletrodeposited membranes with pore diameters of D = 18 nm and a pore center-to-center distance of d = 56 nm, which we call NiFe-A (permalloy), Fe (iron), and Co (cobalt). We manufacture another sample, i. e, a eletrodeposited membrane with a pore diameter of D = 44 nm and a pore center-to-center distance of d = 98 nm, which we call NiFe-B (permalloy). All eletrodeposited nanowires have a length on the order of 1 to 5 $\mu$m.

In \textbf{Figs. 2 (a)}, \textbf{(b)}, \textbf{(c)} and \textbf{(d)} we show the maps of experimental interactions in $\theta_H = 00^{\circ}$,
\begin{figure}[h]
	\vspace{0.1mm} \hspace{0.1mm}
	\begin{center}
		\includegraphics[scale=0.26]{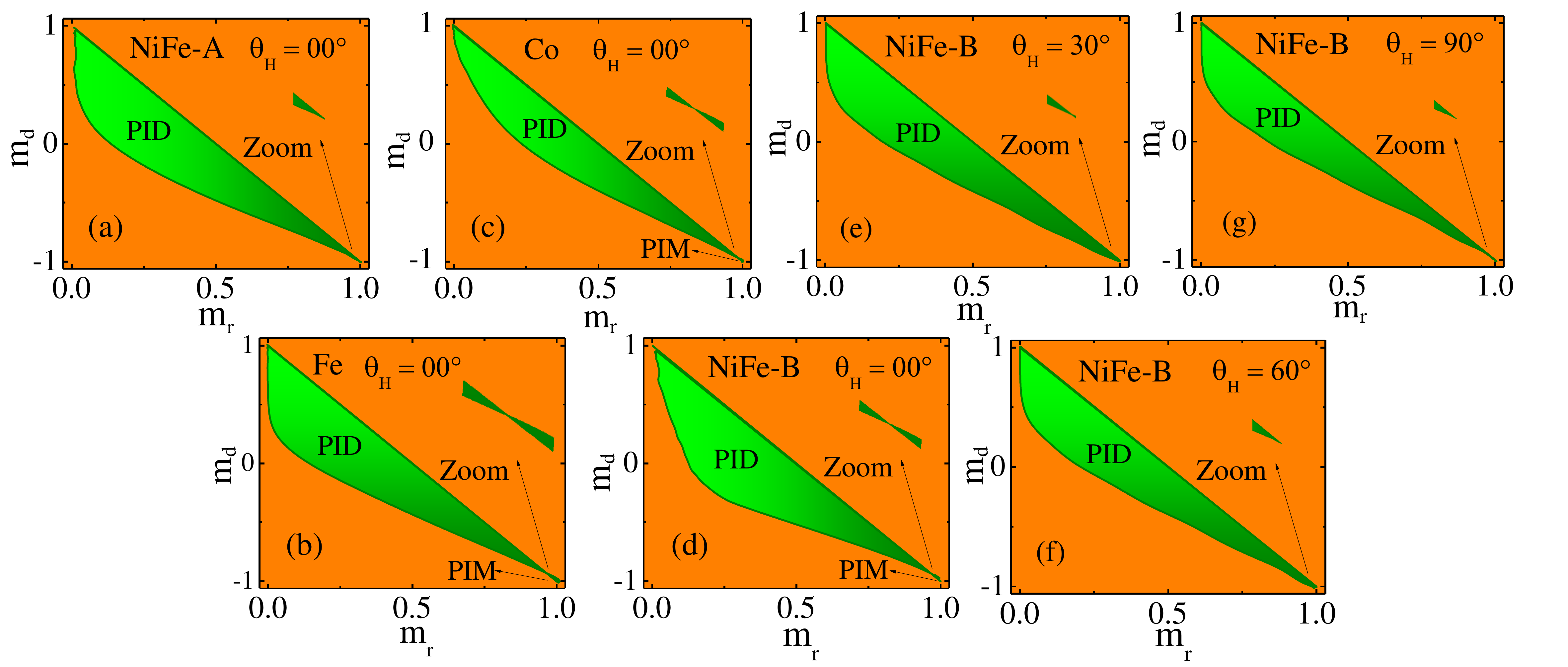}
		\caption{\label{arttype}(Color online) (a) NiFe-A, (b) Fe, (c) Co, and (d) NiFe-B show the experimental interactions maps at $\theta _{H} = 0^{\circ}$ for the different materials. (e) $\theta_{H} = 30^{\circ}$, (f) $\theta_{H} = 60^{\circ}$ and (g) $\theta_{H} = 90^{\circ}$ show the angular dependence of experimental interactions maps for the NiFe-B sample, which has the largest amount of electrodeposited magnetic material.}
		\label{puga}
	\end{center}
\end{figure}
where the hatched areas represent the intersection of the curves defined by the equations
\begin{equation} 
	m_{d2} (H) = \Delta m_{Exp} + [1 - 2m_{r} (H)]
	\label{3}
\end{equation} 
and
\begin{equation} 
	m_{d1} (H) = 1 - 2m_{r} (H).
	\label{3}
\end{equation}
Here $m_{d2} = DCD(H)_{2}/IRM(H_{Max})$, $m_{d1} = DCD(H)_{1}/IRM(H_{Max})$ and $m_{r} = IRM(H)/IRM(H_{Max})$ are the direct current demagnetizations (DCD(H)) and isothermal remanent magnetization (IRM(H)) curves, all normalized by the maximum value of the isothermal remanent magnetization (IRM(H)) curve for the maximum magnetic field H = H$_{Max}$. The experimental term $\Delta m_{Exp}$ represents the magnetic interactions. In the \textbf{Fig. 2 (a)}, we observe only PID interactions, which convey the influence of the amount of electrodeposited magnetic material. On the other hand, in the experimental interaction maps presented in \textbf{Figs. 2 (b)-(d)}, we illustrate the evidence of PID and PIM interactions for each material (iron, cobalt, and permalloy). In the \textbf{Table 1} we present the intensity values found for the PID and PIM interactions using the equations (1) and (2) in $I = lim_{||\delta|| \rightarrow 0} \sum_{i = 1}^{N} |\Delta m_{Exp, i}|\delta_{i}m_{r} (H)$ [27], where $\Delta m_{Exp, i} < 0$ for PID and $\Delta m_{Exp, i} > 0$ for PIM. The values show the behavior from the magnetic interactions in each sample defining an intrinsic characteristic that can be used to define the behavior of interactions in devices.

\begin{table}[h]
	\caption{\label{tab:table4}{Show intensity values from the magnetic interactions of the PID and PIM interactions for each sample (NiFe-A, Fe, Co, and NiFe-B) at $\theta_{H} = 0^{\circ}$.}}%
	\begin{center}
	\begin{tabular}{ccc}
		& PID & PIM \\
		\hline
		NiFe-A & $0.43 \pm 0.003$ & 0 \\
		Fe & $0.41 \pm 0.004$  & $0.00093 \pm 0.000002$  \\
		Co & $0.32 \pm 0.003$  & $0.00018 \pm 0.000005$  \\
		NiFe-B & $0.43 \pm 0.004$ & $0.00029 \pm 0.000004$  \\
	\end{tabular} 
\end{center}
\end{table}

In the \textbf{Figs. 2 (d)}, \textbf{(e)}, \textbf{(f)} and \textbf{(g)} we show the behavior of the angular dependence of the experimental interaction maps of the NiFe-B sample for $\theta_H = 00^{\circ}$, $\theta_H = 30^{\circ}$ , $60^{\circ}$ and $90^{\circ}$. These maps show the variation in the intensity of the experimental term $\Delta m_{Exp}$. We observe from the experimental interaction maps of the \textbf{Figs. 2 (d)}, \textbf{(e)}, \textbf{(f)} and \textbf{(g)}, that the PID and PIM interactions decrease as the angle $\theta_H$ increases. Such behavior is a result of the decrease in energy balance during the IRM and DCD measurements. 

The mean-field approach introduced by Ref. [13] is an estimation of the dipolar energy in real nanowire arrays, which is an aspect important in this type of sample. Even in electrodeposited membranes with equally long nanowires, the dipole field is not uniform along the length of the nanowire. The inhomogeneous dipolar field leads to dipolar interactions between nanowires; the situation is even more complicated here because the length of the nanowire is not uniform. In general, the analysis of the angular dependence of the PID interactions for nanowire arrays needs to consider the energy density of packing of the nanowires in the array [13], which  may  be  written  as $E_{PID}^{NWA} = -3 \pi P M^{2} sin^2 (\varphi_{PID}+\theta_H)$, where $P = (\pi/2\sqrt{3})(D/d)^2$ is the packing factor of the nanowires in the array, d is the interwire distance and D the diameter of nanowire and $\varphi_{PID}$ is the PID angle defined by $\varphi_{PID} = \theta_{PID} - \theta_H$. In this way, the angular dependence of intensity value for the PID interactions is described here for nanowire arrays by
\begin{equation} 
	I_{PID} = I_{PID}^{NW}(0) cos(\theta_{H}) + I_{PID}^{NWA}(0) sin^{2}(\varphi_{PID} + \theta_{H}),
	\label{4}
\end{equation}
where $I_{PID}^{NW}(0)$ and $I_{PID}^{NWA}(0)$ refer to the maximum values obtained for the intensities of interactions for the individual nanowires (NW) and the nanowire arrays (NWA) [26]. In the \textbf{Fig. 3} we present the angular variation of the PID and PIM interactions of the NiFe-B sample obtained through the experimental interaction maps using the same procedure realized to obtain the values in \textbf{Table 1}. We chose to analyze this sample because it has the largest amount of electrodeposited magnetic material. Fitting the data of the \textbf{Fig. 3 (a)} with Equation (3), we found $I_{PID}^{NW}(0) = 0.42$, $I_{PID}^{NWA}(0) = 0.35$ and $\varphi_{PID} = -7^{\circ}$. The values obtained for the intensities $I_{PID}^{NW}(0)$ and $I_{PID}^{NWA}(0)$ are in good agreement with the values of the \textbf{Table 1} for PID interactions. 

According to \textbf{Fig. 3 (b)}, the NiFe-B sample exhibits PIM interactions. The PIM interactions are associated with the exchange interaction between grains or nanowires. The exchange energy density can be considered constant in the nanowires and for nanowire arrays can be written as $E_{Exc} = - A cos (\varphi_{PIM} + \theta_{H})$, where $A$ is the exchange constant [3, 7, 8, 11, 12, 25] and $\varphi_{PIM}$ is the PIM angle defined by $\varphi_{PIM} = \theta_{PIM} - \theta_H$. Hence, the angular dependence of intensity value for the PIM interactions is written as [26]:
\begin{equation} 
	I_{PIM} = I_{PIM}^{NW} (0) + I_{PIM}^{NWA} (0) cos(\varphi_{PIM}+\theta_{H}),
	\label{4}
\end{equation}
where $I_{PIM}^{NW}(0)$ and $I_{PIM}^{NWA}(0)$ represent the maximum value found for the intensities of interactions for the nanowires (NW) and nanowire arrays (NWA). Considering the data of the \textbf{Fig. 3 (b)} and its respective fit with Equation (4), we obtained $I_{PIM}^{NW}(0) = 0.00049$, $I_{PIM}^{NWA}(0) = 0.00052$ and $\varphi_{PIM} = 114^{\circ}$.  
\begin{figure}[h]
	\vspace{0.1mm} \hspace{0.1mm}
	\begin{center}
		\includegraphics[scale=0.36]{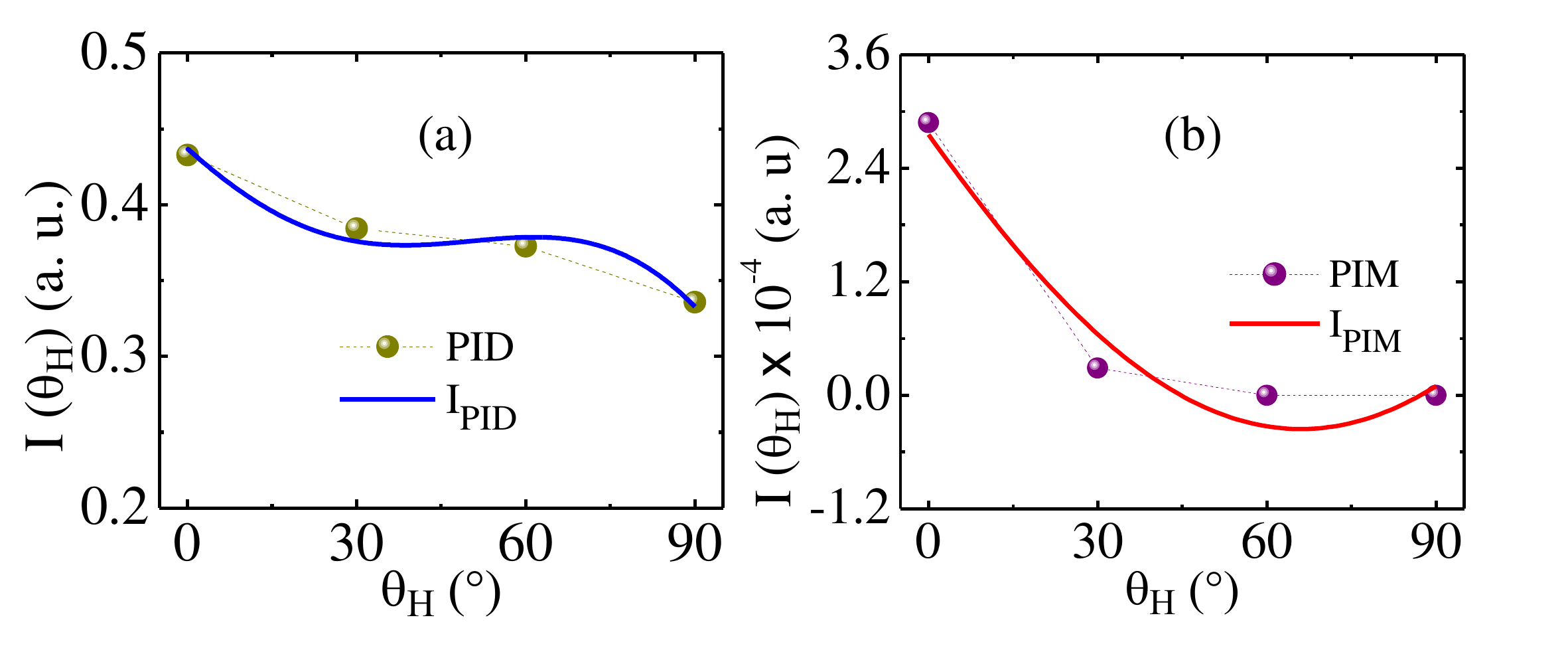}
		\caption{\label{arttype}(Color online) Shows the angular dependence from the magnetic interactions PID (a) and PIM (b) for the NiFe-B sample.}
		\label{puga}
	\end{center}
\end{figure}
The values obtained for the intensities $I_{PIM}^{NW}(0)$ and $I_{PIM}^{NWA} (0)$ are in perfect agreement with \textbf{Table 1} for PIM interactions. The angles $\varphi_{PID} = -7^{\circ}$ and $\varphi_{PIM} = 114^{\circ}$ reveal the positions of maximum equilibrium of the spins and its liquid magnetic moment associated during the experimental measurements. We observed a $\varphi = -7^{\circ} + 114^{\circ} = 107^{\circ} $ phase difference between the maximum equilibrium positions of the PID and PIM interactions. The observed phase difference is unequivocal evidence that we can separate magnetic interactions. Furthermore, the results show that magnetic interactions in nanostructures are complex (depending on material, shape, experimental configuration, and excitation parameters), and its analysis is fundamentally necessary for applications. The research supported by the experimental results presented here does not converge to the dependence of any magnetic instabilities in applications.

Nanostructures, such as nanowires, are used in a diversity of applications spreading from nanoscale optoelectronic devices (such as sensors, for example) to high-density data storage media. The remnant state offers a fundamental and natural way of studying magnetic interactions for applications. In summary, the approach we present here incorporates the measurement of magnetic interactions experimentally. Furthermore, it represents a way of analyzing the value of the intensity of interactions considering them as PID and PIM, which is of fundamental importance to describe the behavior of magnetic interactions in devices.

\section*{Conflicts of interest}

There are no conflicts of interest to declare.

\section*{Data availability statement}

Data underlying the results presented in this paper are not publicly available at this time but may be obtained from the authors upon reasonable request.

\section*{Acknowledgments}

This research was supported by the Conselho Nacional de Desenvolvimento Científico e Tecnológico (CNPq), Coordenação de Aperfeiçoamento de Pessoal de Nível Superior - Universidade Federal Rural de Pernambuco (CAPES-UFRPE), Financiadora de Estudos e Projetos (FINEP) and Fundação de Amparo a Ciência e Tecnologia de Pernambuco (FACEPE). Adrielson de Araújo Dias acknowledges the Program for Scientific Initiation Scholarships (PIBIC) of the CNPq/UFRPE. 

\bibliographystyle{MiKTeX}

\end{document}